\documentclass[aps,prf,amsmath,amssymb,aps,preprint,groupedaddress]{revtex4-2}

\usepackage{graphicx}

\begin{document}

\newcommand{\U}{{\bf U}}
\newcommand{\bA}{{\bf A}}
\newcommand{\bB}{{\bf B}}
\newcommand{\bG}{{\bf G}}
\newcommand{\bI}{{\bf I}}
\newcommand\bc{\boldsymbol c}
\newcommand\bh{\boldsymbol h}
\newcommand\bff{\boldsymbol f}
\newcommand\bg{\boldsymbol g}
\newcommand\bL{\boldsymbol L}
\newcommand\bM{\boldsymbol M}
\newcommand\bp{\boldsymbol p}
\newcommand\bP{\boldsymbol P}
\newcommand\bPi{\boldsymbol \Pi}
\newcommand\bLambda{\boldsymbol \Lambda}

\newcommand\ba{\boldsymbol a}
\newcommand\bb{\boldsymbol b}
\newcommand\bn{\boldsymbol n}
\newcommand\bk{\boldsymbol k}
\newcommand\bx{\boldsymbol x}
\newcommand\by{\boldsymbol y}
\newcommand\bz{\boldsymbol z}
\newcommand\bq{\boldsymbol q}
\newcommand\bQ{\boldsymbol Q}
\newcommand\bH{\boldsymbol H}
\newcommand\bu{\boldsymbol u}
\newcommand\bU{\boldsymbol U}
\newcommand\bv{\boldsymbol v}
\newcommand\bV{\boldsymbol V}
\newcommand\bw{\boldsymbol w}
\newcommand\bW{\boldsymbol W}
\newcommand\be{\boldsymbol e}
\newcommand\br{\boldsymbol r}
\newcommand\bt{\boldsymbol t}
\newcommand\bo{\boldsymbol 0}
\newcommand\boo{\boldsymbol o}
\newcommand\bOO{\boldsymbol O}
\newcommand\btau{\boldsymbol \tau}
\newcommand\bC{\boldsymbol C}

\newcommand\bS{\boldsymbol S}
\newcommand\bT{\boldsymbol T}
\newcommand\bcalG{\boldsymbol{\cal G}}
\newcommand\bcalT{\boldsymbol{\cal T}}
\newcommand\bcalL{\boldsymbol{\cal L}}
\newcommand\bcalM{\boldsymbol{\cal M}}
\newcommand\bcalQ{\boldsymbol{\cal Q}}
\newcommand\bcalR{\boldsymbol{\cal R}}
\newcommand\bcalS{\boldsymbol{\cal S}}
\newcommand\bcalA{\boldsymbol{\cal A}}
\newcommand\bcalB{\boldsymbol{\cal B}}
\newcommand\bcalC{\boldsymbol{\cal C}}
\newcommand\bcalD{\boldsymbol{\cal D}}
\newcommand\bPhi{\boldsymbol{\Phi}}
\newcommand\bSigma{\boldsymbol{\Sigma}}
\newcommand\bvarphi{\boldsymbol{\varphi}}
\newcommand\bOmega{\boldsymbol{\Omega}}
\newcommand\bChi{\boldsymbol{\chi}}
\newcommand\bE{\boldsymbol{E}}
\newcommand\bF{\boldsymbol{F}}
\newcommand\bR{\boldsymbol{R}}
\newcommand\bbf{\boldsymbol{f}}
\newcommand\ip{\boldsymbol\cdot}
\newcommand\dip{\boldsymbol :}

\title{Hierarchical Higher-Order Dynamic Mode Decomposition for Clustering and Feature Selection}


\author{Adrián Corrochano}
	\email{adrian.corrochanoc@upm.es}
	\affiliation{School of Aerospace Engineering, Universidad Politécnica de Madrid, 28040 Madrid, Spain}
\author{Giuseppe D'Alessio}
	\affiliation{Universit\'{e} Libre de Bruxelles, \'{E}cole polytechnique de Bruxelles, Aero-Thermo-Mechanics Laboratory, Bruxelles, Belgium}
	\affiliation{CRECK Modeling Lab, Department of Chemistry, Materials and Chemical Engineering, Politecnico di Milano, Piazza Leonardo da Vinci 32, 20133 Milano, Italy}
\author{Alessandro Parente}
	\affiliation{Universit\'{e} Libre de Bruxelles, \'{E}cole polytechnique de Bruxelles, Aero-Thermo-Mechanics Laboratory, Bruxelles, Belgium}
\author{Soledad Le Clainche}
	\affiliation{School of Aerospace Engineering, Universidad Politécnica de Madrid, 28040 Madrid, Spain}

\date{\today}

\begin{abstract}
In this work, a new algorithm based on the application of higher-order dynamic mode decomposition (HODMD) is proposed for feature selection and variables clustering in reacting flow simulations. 
The hierarchical HODMD (h-HODMD) performs a reduction of the model order, followed by the iterative selection of the best reconstructed variables thus creating clusters of features which can eventually be associated with distinct dynamical phenomena. 
Firstly, h-HODMD is combined with different data pre-processing techniques to assess their influence on the algorithm in terms of reconstruction error. 
Afterwards, the proposed algorithm is applied to analyze three different databases obtained from numerical simulations of a non-premixed co-flow methane flame, and its performance are compared with the standard HODMD in terms of the achievable degree of reduction as well as in terms of reconstruction error. 
Two detailed kinetic mechanisms are used for the simulations, and different time horizons are considered to assess the influence of the number of considered cycles on the accuracy of the reduced model. 
Results show that h-HODMD improves the reconstruction for all the variables when compared to the standard HODMD algorithm. 
This condition is achieved thanks to the iterative variables' clustering: finding dedicated modes for a specific group of features does in fact lead to a better reconstruction of the dynamics with respect to the case when the same (global) modes are used to reconstruct the entire set of variables.
Finally, the clusters of variables found by means of h-HODMD are analyzed, and it is observed that the algorithm can group, in an unsupervised fashion, chemical species whose behavior is also consistent from a kinetic point of view.
In fact, radicals responsible for fast chemistry processes (such as propagation and branching reactions) are grouped together, and the same happens for nitrogen oxides (whose chemical timescales are instead order of magnitudes slower). 
The possibility to identify and to describe correctly the main dynamics with reduced order models for these two classes of chemical species is extremely important for practical applications in combustion.
In fact, it allows for the possibility to formulate inexpensive reduced dynamical models for predicting flames liftoff, as well as for identifying the formation of local extinction and blowout conditions, to formulate accurate reduced models to describe the formation of pollutants in aviation, and for control purposes.
\end{abstract}


\maketitle

\section{Introduction \label{sec:Intro}}

Understanding the complex physical and chemical phenomena occurring in reacting systems is crucial because of the urgent need for a reduction of the environmental impact of the next generation combustion systems. 
However, performing detailed experimental and/or numerical investigations of combustion systems (especially when large-scale systems are taken into account) is not always straightforward because of the associated high computational cost  \cite{cant2002high}. 

Thus, the adoption of reduced order models (ROMs) represents a viable option in this context, as the latter are capable of describing complex reacting systems with low errors and minimal computational costs \cite{parente2011investigation, bellemans2018feature, Coussement2013}. 

One of the most employed algorithms to construct ROMs for reacting flows is represented by the principal component analysis (PCA) \cite{Jolliffe2011}, also known as proper orthogonal decomposition (POD) \cite{berkooz1993proper}.
The latter is a dimensionality reduction algorithm based on the projection of the original system onto a lower-dimensional manifold whose directions (the principal components, PCs) are orthogonal and obtained as linear combination of the original variables.
In literature, it is possible to find many applications exploiting PCA to speed-up computational fluid dynamics (CFD) simulations of reacting systems by reducing the number of the differential equations to be solved for combustion or atmospheric re-entry applications \cite{isaac2014reduced, malik2021combustion,bellemans2017reduced,d2022automated}.
In addition, the algorithm and its local formulation were recently validated for clustering in chemical kinetics applications \cite{d2020adaptive, d2020impact}, for data analysis of Direct Numerical Simulation (DNS) of reacting jets \cite{d2020analysis,d2020unsupervised}, and for feature extraction and selection to enhance the classification and regression accuracy of Artificial Neural Networks (ANN) \cite{tipler2022predicting,d2021feature}.

Recently, dynamic mode decomposition (DMD) \cite{Schmid10} was proposed for the analysis and the development of ROMs of unsteady systems.
The main advantage for this algorithm over alternative modal decomposition techniques, such as PCA, is represented by the possibility to retrieve the frequency associated to the modes to analyze the system. 
In addition, DMD finds patterns that can be related to flow instabilities \cite{Souvick2013,Quinlan2014,Huang2016,Motheau2014}, and it can also be used for dimensionality reduction, as fewer DMD modes are usually capable of reproducing the main flow dynamics with a reduced computational cost. 
Nevertheless, its applicability is usually limited to specific conditions \cite{LeClaincheVegaComplexity18}. 
Thus, higher order dynamic mode decomposition (HODMD) \cite{LeClaincheVega17} was found to be a more robust alternative to standard DMD for non-reactive complex flows and noisy experimental data, and it has been widely used in the analysis of several complex flows in industrial applications, either to create ROMs or to study flow patterns \cite{Corrochano,Lazpita2022,LeClaincheetalJAircraft18,LeClaincheetalJFM2020}. 
Recently, HODMD was also proposed and validated for the development of reduced order models for reacting flows \cite{Corrochano22}.

In the current work, a modified formulation for HODMD is proposed to accomplish feature selection and clustering of variables when high-dimensional reacting systems are considered. More specifically, the hierarchical HODMD (hereinafter referred to as h-HODMD), iteratively identifies the best reconstructed variables and associates them with the corresponding DMD modes and frequencies. 
The performance of the proposed algorithm are assessed using three databases obtained from CFD simulations of a laminar co-flow flame taking into account different kinetic mechanisms, as well as different boundary conditions.
The two kinetic mechanisms used to describe methane oxidation consist of 84 and 53 chemical species, respectively.
A wide range of chemical time scales is also considered since the nitrogen oxides' chemistry is also included in one of the two mechanisms.
The considered flame, despite being laminar, exhibits complex physics because of local extinction phenomena.
Finally, the results are compared with the ones obtained by using a standard HODMD in terms of the achievable degree of reduction as well as in terms of reconstruction error.

The current work is organized as follows. 
In Section \ref{sec:Meth}, the original HODMD is explained and the modifications to obtain the h-HODMD are discussed. 
In Section \ref{Num}, the numerical setup for the CFD simulations that were considered for the training datasets is shown and discussed. 
Finally, the main results and conclusions are presented in Sections \ref{Results} and \ref{sec:conclusions}, respectively.

\section{Higher Order Dynamic Mode Decomposition \label{sec:Meth}}

Higher Order Dynamic Mode Decomposition (HODMD) \cite{LeClaincheVega17} is a data driven method used in fluid dynamics to identify flow patterns. 
The latter represents an extension of Dynamic Mode Decomposition (DMD) \cite{Schmid10} used for the analysis of complex flow \cite{LeClainche19}, noisy experiments \cite{Mendez21} or turbulent flows \cite{LeClaincheHanFerrer19}. 
HODMD decomposes data equispaced in time as an expansion of Fourier-like DMD modes as

\begin{equation}
	\bv(x,y,t_{k})\simeq  \sum_{m=1}^M a_{m}\bu_m(x,y)e^{(\delta_m+i \omega_m)t_k},\quad
	k=1,\ldots,K,\label{ab00}
\end{equation}
where $\bu_m$ are the DMD modes and $a_{m}$, $\omega_m$, $\delta_m$ are their associated amplitudes, frequencies and growth rates. 
The number of DMD modes ($M$) is referred to as {\it spectral complexity}, whereas $K$ in Eq. \ref{ab00} represents the number of snapshots available for the analyzed data, which is referred to as the {\it temporal dimension}. 
A thorough description of the algorithm can be found in Ref. \cite{LeClaincheVega17}, while the associated code is available in Ref. \cite{HODMDbook}.

\subsection{Main algorithm \label{sec:Main}}

Firstly, equidistant snapshots in time for the considered dynamical system (with time interval $\Delta t$) are organized as the following snapshot matrix
\begin{equation}
	\bV_1^K = [\bv_{1},\bv_{2},\ldots,\bv_{k},\bv_{k+1},\ldots,\bv_{K-1},\bv_{K}].\label{ab0}
\end{equation}
From the definition in Eq. \ref{ab0}, it follows that the dimensions of the input matrix are thus $J \times K$, where $K$ is the number of snapshots analyzed and $J = N_x \times N_y$ is the number of grid points considered for the CFD simulations. 
The HODMD algorithm can be summarized into two main steps: first, a spatial reduction via singular value decomposition (SVD) is accomplished, then the DMD-d algorithm is applied.

Within the first step, SVD is applied to the snapshot matrix to reduce noise and spatial redundancies to only retain the large scales:

\begin{equation}
	\bV_1^K = \bU \bSigma \bT^T = \bU \hat{\bV}_1^K,
\end{equation}

where $\hat{\bV}_1^K$ is the {\it reduced snapshot matrix} whose dimensions are $N \times K$. 
The number of SVD modes to be retained ($N$) can be determined setting a tolerance $\varepsilon_1$ as 
\begin{equation}
	\dfrac{\sigma_{N+1}}{\sigma_1}\leq \varepsilon_1.\label{e26}
\end{equation}
By means of the operations discussed above, the spatial complexity can be reduced from $J$ grid points to $N$ linearly independent vectors.
When high-dimensional data are analyzed, such as the ones obtained from complex reacting systems, SVD is replaced with Higher Order Singular Value Decomposition (HOSVD) \cite{HOSVD}. 
To perform this decomposition, the snapshot matrix $\bV_1^K$ is organized in tensor form $\bV_{ijlk}$ having dimensions $N_v \times N_x \times N_y \times K$. 
HOSVD can then be used to reduce the dimensionality in an efficient way. 
The algorithm applies standard SVD to the four matrices whose columns are formed by each one of the four dimensions of the tensor. 
In particular, if a four-dimensional tensor is considered, HOSVD is applied as follows

\begin{equation}
	\bV_{ijlk}\simeq\sum_{p_1=1}^{N_v}\sum_{p_2=1}^{N_x}\sum_{p_3=1}^{N_y}\sum_{n=1}^{K} \bS_{p_1p_2p_3n}
	\bW^{(v)}_{ip_1}\bW^{(x)}_{jp_2} \bW^{(y)}_{lp_3} \bT_{kn},    \label{c10}
\end{equation}

where $S_{p_1p_2p_3n}$ is the core tensor and $\bW^{(v)}$, $\bW^{(x)}$, $\bW^{(y)}$ and $\bT$ are the modes of the decomposition in each dimension. As well as in SVD, it is possible to retain a number of modes en each direction, in order to remove the noise and retain the large scales.
The temporal modes matrix $\bT$ is used as the reduced snapshot matrix.

In the second step, i.e., the application of the {\it DMD-d algorithm}, the high order Koopman assumption is applied to the reduced snapshot matrix as follows	

\begin{equation}
	\hat\bV_{d+1}^K\simeq \hat\bR_1 \hat\bV_1^{K-d}+ \hat\bR_2 \hat\bV_2^{K-d+1} + \ldots + \hat\bR_d \hat\bV_d^{K-1}.\label{ab6}
\end{equation}

The Koopman operators $\hat\bR_1$, $\hat\bR_2$, $\cdots$, $\hat\bR_d$ are encompassed into a single matrix, which contains the dynamics of the system. 
When $d=1$ is considered, the algorithm provides the same results to the standard DMD.

The eigenvalues provide the frequencies $\omega_m$ and the growth rates $\delta_m$ of the eigenvectors, which are used to construct the DMD modes $\bu_m$. 
Finally, the amplitude of the DMD modes $a_m$ are obtained via least squares fitting of the expansion which was previously expressed in Eq. \ref{ab00}. 
Using a second tunable tolerance $\varepsilon_2$, it is possible to then select $M$ DMD modes that are representative for the main dynamics of the system as follows

\begin{equation}
	a_{M+1}/a_1\leq \varepsilon_2. \label{b66}
\end{equation}

When the considered database is related to the transient state of a numerical simulation, it is necessary to perform another step to identify the leading modes driving the flow dynamics. This step consists in computing the contribution to the transient state $I_m$ defined as
\begin{equation}
	I_m = \sum_{k = 1}^{K} \lvert a_m e^{(\delta_m + i\omega_m)t_k}\rvert \;\|u_m\|_{F}^{2} \times \Delta t, \label{e42}
\end{equation}
where $\| \bullet \|_{F}^2 $ is the square of the Frobenius norm. 
This index $I_m$ takes into account also the damping or growth of the modes in time, which is reflected in $\delta_m$. 
A much more detailed description about this criterion can be found in \cite{Kou}.

Finally, the accuracy of the algorithm is measured in terms of the relative root mean square error (RRMSE) as

\begin{equation}
	\text{RRMSE}=\sqrt{\frac{\|\bV^\text{DMD}-\bV\|_F}{\|\bV\|_F}}.\label{error}
\end{equation}

\subsection{Pre-processing techniques for applying HODMD to combustion databases \label{sec:scaling}}

When HODMD is applied to combustion databases, a pre-processing step of the data in terms of centering and scaling must be applied because of their multivariate nature \cite{PCA, Corrochano22}.
Centering is accomplished by subtracting the temporal mean of the considered variable to each statistical observation of the dataset, so that each sample can be seen as a fluctuation from the mean. 
Scaling the variables, instead, is motivated by the need of having them lying in the same order of magnitude. 
The centered and scaled $i$-th statistical observation for the $j$-th variable is finally obtained as
\begin{equation}
	\tilde{x}_{i,j} = \dfrac{x_{i,j}-\bar{x}_{i,j}}{c_j}\label{e24},
\end{equation}
\noindent where $\bar{x}_{i,j}$ represents the temporal mean of the $i$-th point considering the variable's vector $x_j$ and $c_j$ is the associated scaling factor. 
Four scaling techniques were considered for the current analysis:

\begin{enumerate}
	\item \textit{Range} scaling: The difference between the maximum and minimum value of each variable is used as scaling factor.
	\item \textit{Auto} scaling: the standard deviation of the $j$-th variable ($\sigma_{j}$) is used as scaling factor.
	\item \textit{Pareto} scaling: the square root of the standard deviation of the $j$-th variable ($\sqrt{\sigma_{j}}$) is used as scaling factor.
	\item \textit{Vast} scaling: the product of the standard deviation $\sigma_{j}$ and the coefficient of variation $\sigma_{j} / \bar{x}_{j}$ is used as scaling factor. 
\end{enumerate}

The effects of the different scaling factors on modal decomposition techniques such as PCA and HODMD have been throughly analyzed for combustion applications in literature \cite{PCA,d2020analysis,Corrochano22}. 

\subsection{Hierarchical Higher Order Dynamic Mode Decomposition\label{sec:HHODMD}}

The hierarchical higher order dynamic mode decomposition (h-HODMD) is a new algorithm which aims at iteratively identifying the main dynamics of the considered system, creating clusters of variables having the same dynamic behavior.

After applying the chosen pre-processing technique to the considered database, HODMD is applied to the scaled tensor. 
Once the model's order is reduced, each variable can be reconstructed back to the original space by multiplying by the scaling factor and adding the mean value, as

\begin{equation}
	\bx_j^{DMD} = \tilde{\bx}_j^{DMD}*c_j+\bar{\bx}_j, \label{e25}
\end{equation}

The accuracy of the reconstruction of each variable is measured afterwards by using the relative root mean square error. 
A tolerance $\varepsilon_v$ is selected by the user to retain the variables verifying  the criterion

\begin{equation}
	\sqrt{\frac{\|\bx_j^\text{DMD}-\bx_j\|_F}{\|\bx_j\|_F}}<\varepsilon_v.\label{errorV}
\end{equation}

\begin{figure}
	\centering
	\includegraphics[width = \linewidth]{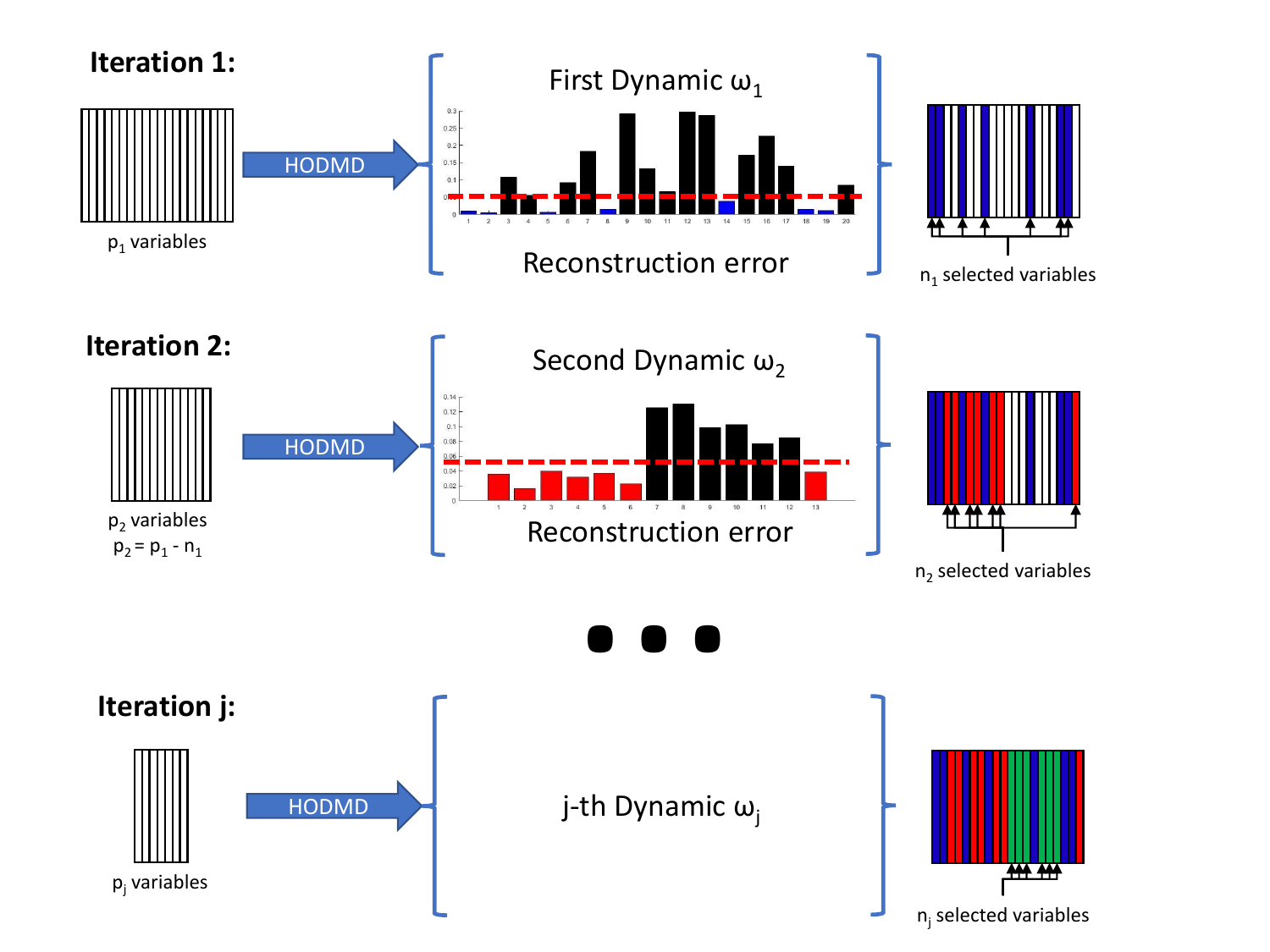}
	\caption{Sketch for the h-HODMD algorithm: the variables of the considered dynamical system are progressively grouped together according to their reconstruction error, and clusters are obtained in an unsupervised fashion. \label{fig:HHODMD}}
\end{figure}

Afterwards, HODMD is applied again to the remaining variables so they can be grouped according to their reconstruction error. The variables of the considered system are progressively better reconstructed as HODMD is applied at each iteration, and new modes and frequencies are obtained. 
A summary of the iterative process is shown in Figure \ref{fig:HHODMD}: a group of variables (cluster) is retained at each iteration and it is removed from the dataset for the next iteration of HODMD. 
In this work, the number of clusters created in each analysis (i.e., the number of iterations for the algorithm) is set to $4$.

\section{Numerical simulations \label{Num}}

The CFD simulations carried out to obtain the databases analyzed by h-HODMD in the following sections are representative for the oxidation of a nitrogen-diluted methane jet ($65\%$ of methane and $35\%$ nitrogen on a molar basis) in air.
The oxidizer is injected into the domain with a constant velocity of 35 $cm/s$, while the fuel is injected according to the following sinusoidal velocity profile:

\begin{equation}
	v(r,t) = v_{max} \left( 1 - \frac{r^2}{R^2} \right) [1 + A \ sin(2 \pi f \ t)],
\end{equation}

where $v_{max}$ is the maximum velocity ($v_{max} = 70 cm/s$), $r$ is the radial coordinate, $R$ represents the internal radius of the nozzle and $A$ is the amplitude of the perturbation ($A = 0.25$). 
\par
Three numerical simulations were carried out changing the frequency of the perturbation $f$ and/or the kinetic mechanism, as summarized in Table \ref{tab:Freq}. 
The two detailed kinetic mechanisms employed for the CFD simulations were the \texttt{POLIMI\_C1\_C3\_HT\_1412} \cite{POLI}, which takes into account 84 species and 1698 reactions, and the \texttt{GRI-Mech\ 3.0} \cite{smith1999gri}, which instead consists of 53 species and 325 reactions.
The simulations carried out using the \texttt{GRI-Mech\ 3.0} represent a particularly interesting test case for the proposed algorithm. 
In fact, the kinetic mechanism takes into account the chemistry associated to the formation and the consumption of nitrogen oxides (NO$_{x}$), and both phenomena are well-known to take place on completely different timescales with respect to other chemical species and radicals.
Thus, this represents an additional degree of complexity for the formulation of an accurate reduced-order model for the dynamical systems being considered in this work.

The CFD simulations were carried out by using \texttt{LaminarSMOKE}, an \texttt{OpenFOAM}-based operator-splitting solver proposed by Cuoci \emph{et al.} \cite{cuoci2013numerical}. 
Additional information regarding the numerical settings for the simulations can be found in Refs. \cite{d2020adaptive,d2020impact}.

\begin{table}
	\centering
	\begin{tabular}{|c|c|c|c|}
		\hline
		& frequency [Hz] & Number of species & Number of reactions  \\\hline
		Simulation 1 & 10	& 84 & 1698\\
		\hline
		Simulation 2 & 10	&  53 & 325\\
		\hline
		Simulation 3 & 20	& 53 & 325\\
		\hline
	\end{tabular}
	\caption{Frequency of the perturbation (in Hertz) and number of species and reactions being used for the CFD simulations.\label{tab:Freq}}
\end{table}

For each simulation reported in Table \ref{tab:Freq}, a database is extracted in the form of a fourth-order tensor having dimensions $n_v \times n_x \times n_y \times n_t$, where 
\begin{itemize}
	\setlength\itemsep{0.01em}
	\item $n_v$ is the number of considered variables (temperature and $n_v-1$ chemical species, respectively);
	\item $n_x$ is the number of grid points along the streamwise direction;
	\item $n_y$ is the number of grid points along the spanwise direction;
	\item $n_t$ is the number of snapshots equispaced in time, considering $\Delta t = 10^{-3}$.
\end{itemize}

For all considered simulations it was imposed $n_x = n_y = 150$, while $n_v$ was changing depending on the number of species originally taken into account by the kinetic mechanism, and $n_t$ was also different for each database. 
The selected values for $n_v$ and $n_t$ for the three simulations are reported in Table \ref{tab:nt} with the associated number of completed periods (a period is defined as: $T = 2\pi/f$). 
The influence of this parameter on the performance of h-HODMD will also be assessed later in Sec. \ref{Results}.

\begin{table}
	\centering
	\begin{tabular}{|c|c|c|c|}
		\hline
		& $n_v$ & $n_t$ & Completed periods \\\hline
		Database 1 & 83	& 250 & 1\\
		\hline
		Database 2 & 53	& 250 & 3\\
		\hline
		Database 3 & 53	& 250 & 5\\
		\hline
	\end{tabular}
	\caption{Number of variables, snapshots, and completed periods for each database considered for the h-HODMD analysis. \label{tab:nt}}
\end{table}

\section{Results\label{Results}}

The algorithm discussed in Sec. \ref{sec:Meth} was finally applied to the data obtained from the simulations summarized in Tab. \ref{tab:Freq}. 
In particular, the databases were firstly preprocessed by means of the centering and scaling operations reported in Sec. \ref{sec:scaling}; afterwards, the algorithm discussed in Sec. \ref{sec:Main} was applied to extract different groups of variables according to their reconstruction error. 
Finally, the dynamics for each group was analyzed.

To quantitatively evaluate the performance of the algorithm proposed in this work, three different metrics to compute the reconstruction error were employed at each iteration:

\begin{enumerate}
	\item The relative root mean square error (RRMSE) just after the application of HODMD:
	\begin{equation}
		RRMSE_{sc}=\sqrt{\frac{\|\tilde{\bV}^\text{DMD}-\tilde{\bV}\|_F}{\|\tilde{\bV}\|_F}},
	\end{equation}
	where $\tilde{\bV}$ is the centered and scaled snapshot matrix and $\tilde{\bV}^\text{DMD}$ is the reconstruction of this snapshot matrix after the application of HODMD.
	\item The RRMSE with the original tensor:
	\begin{equation}
		RRMSE_{or}=\sqrt{\frac{\|\bV^\text{DMD}-\bV\|_F}{\|\bV\|_F}}.
	\end{equation}
	In this metric, each variable of the reconstructed tensor is multiplied by its scaling factor and the mean value is added, as in eq. \ref{e25}.
	\item The average RRMSE value for the variables:
	\begin{equation}
		RRMSE_{mean}=\dfrac{\sum_{j=1}^{n_v}RRMSE_{or}(x_j)}{n_v}
	\end{equation}
\end{enumerate}

The algorithm was calibrated with the aim to ensure its robustness as well as to obtain the lowest reconstruction error, according to the procedure presented in \cite{HODMDbook,Corrochano22}.
The values for $d$ and $\varepsilon=\varepsilon_1 = \varepsilon_2$ adopted for the current work are reported in Tab. \ref{tab:dVal}.

\begin{table}
	\centering
	\begin{tabular}{|c||c|c|c|c|c|c|c|c|}
		\hline
		&\multicolumn{8}{ |c| }{Scaling Method} \\ \hline \hline
		& \multicolumn{2}{ |c| }{Auto}&\multicolumn{2}{ |c| }{Range}& \multicolumn{2}{ |c| }{Pareto} &\multicolumn{2}{ |c| }{Vast} \\  \hline
		&$d$&$\varepsilon$&$d$&$\varepsilon$&$d$&$\varepsilon$&$d$&$\varepsilon$ \\ \hline
		Database 1 & $6$ & $4 \times 10^{-2}$ & $6$& $6 \times 10^{-4}$ & $6$ & $6 \times 10^{-4}$ & $6$ & $5 \times 10^{-3}$ \\
		\hline
		Database 2 & $15$ & $4 \times 10^{-2}$ & $12$& $6 \times 10^{-4}$ & $12$ & $1 \times 10^{-3}$ & $12$ & $4 \times 10^{-3}$ \\
		\hline
		Database 3 & $12$ & $4 \times 10^{-2}$ & $10$& $6 \times 10^{-4}$ & $12$ & $2 \times 10^{-3}$ & $12$ & $4 \times 10^{-3}$ \\
		\hline
	\end{tabular}
	\caption{Values for $d$ and $\varepsilon$ obtained in the HODMD calibration process for the cases considered in the analysis. \label{tab:dVal}}
\end{table}

For the iterative process in h-HODMD, the tolerance $\varepsilon_v$ (eq. \ref{errorV}) was set equal to $\varepsilon_{v,iter} = 0.02 \times iter$, where $iter$ is the number of iterations accomplished by the algorithm.
Thus, the tolerance becomes progressively higher with the number of iterations. 
The number of variables in each cluster $n_{iter}$ is not kept constant, as this parameter is only controlled by the tolerance $\varepsilon_v$.
Other tolerances, i.e. $\varepsilon_{v,iter} = 0.01 \times iter$ and $\varepsilon_{v,iter} = 0.05 \times iter$, have been used for the calibration. However, a high tolerance means that each iteration retains a high number of variables which can have different dynamics among them. On the other hand, a small tolerance may not include all the variables with the same dynamics in a single cluster.

\begin{figure}
	\centering
	\includegraphics[width = \linewidth]{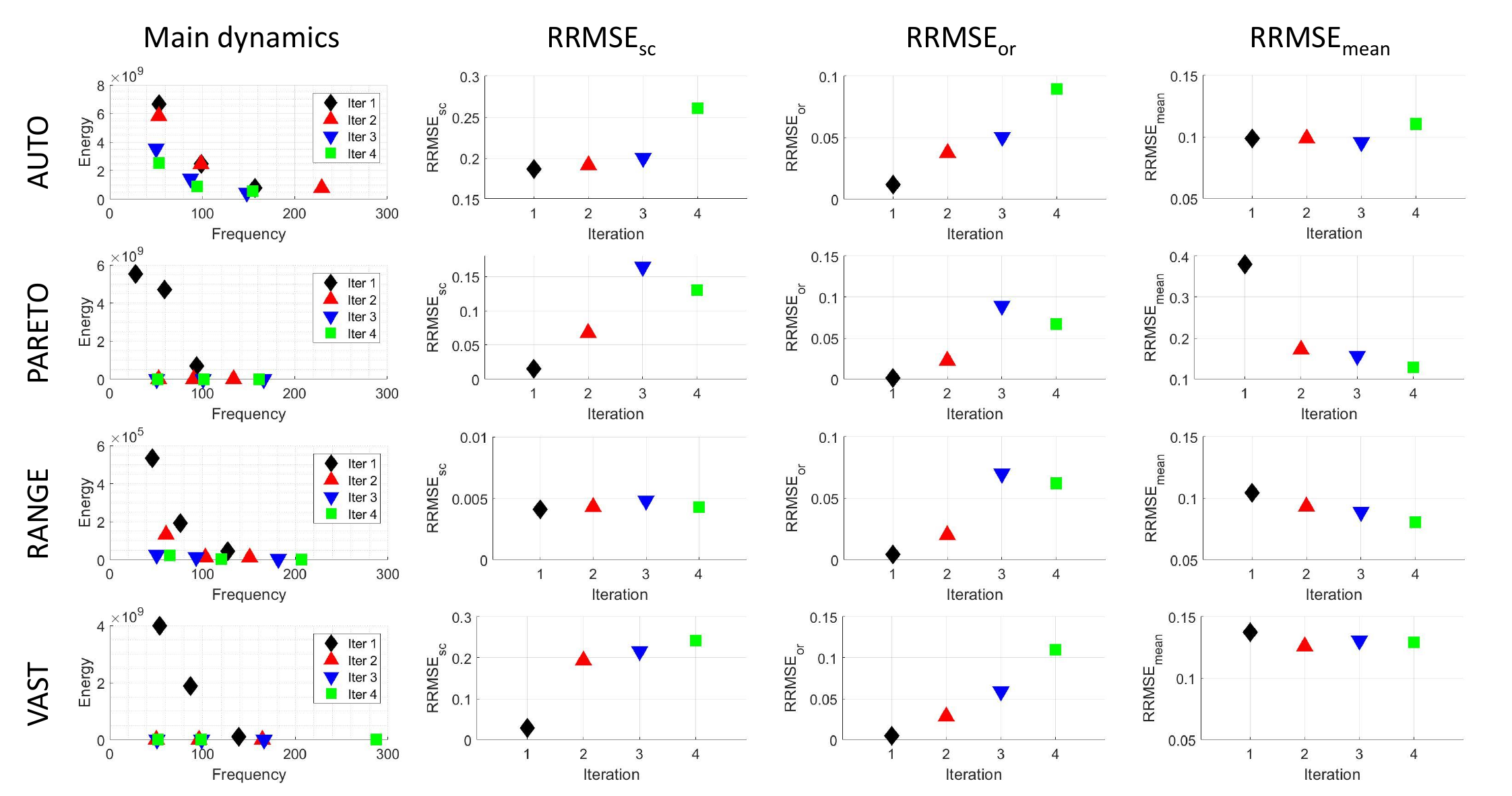}
	\caption{From left to right: Three main frequencies calculated as in Equation \ref{e42}, $RRMSE_{sc}$, $RRMSE_{or}$ and $RRMSE_{mean}$. From top to bottom: auto scaling, range scaling, pareto scaling and vast scaling. The analysis corresponds to Database 1, i.e., to the simulation obtained by imposing a sinusoidal perturbation having frequency $f = 10$ Hz and amplitude A = 0.25, and carried out with a detailed kinetic mechanism accounting for 82 species. \label{fig:Results1}}
\end{figure}

The first database, analyzed in Figure \ref{fig:Results1}, only considers $\sim$ 1 cycle, so the simulation can be considered fully transient. 
Auto scaling appears to provide similar main dynamics at each iteration, and this is in agreement with what is expected from theory (as this scaling method gives the same importance to all the variables). 
Range scaling leads instead to different dynamics for each iteration, while Pareto and Vast scaling initially emphasize temperature and most stable species, respectively, to then give to the rest of variables the same importance as the number of iterations increases. Because of that, after removing the best reconstructed variables, the algorithm give back similar dynamics between two consecutive iterations.

The reconstruction error for the scaled ($RRMSE_{sc}$) and original tensor ($RRMSE_{or}$) increases with the number of iterations, as the algorithm progressively removes the variables which are better reconstructed. The first iteration has a low $RRMSE_{sc}$ for all the scaling methods except auto scaling, as it does not give more relevance to main variables and/or temperature. After the first iteration, the $RRMSE_{sc}$ increases for Pareto and Vast scaling, while it reaches a plateau for Range scaling. 
The qualitative behavior for $RRMSE_{or}$ is similar to the one observed for $RRMSE_{sc}$, with Range scaling providing the best reconstruction error.
Finally, the mean RRMSE ($RRMSE_{mean}$) is stable for auto scaling, and this appears to be justified for it gives to all variables the same importance. 
In range scaling, instead, the $RRMSE_{mean}$ decreases as the number of iterations increases. 
When Pareto scaling is applied, $RRMSE_{mean}$ on the first iteration is much higher than the rest of the iterations. 
This happens because the latter initially emphasizes temperature, thus ignoring the reconstruction of the rest of the variables.
After the first iteration, the rest of the variables become more important, and $RRMSE_{mean}$ decreases. The $RRMSE_{mean}$ when using vast scaling remains almost constant with the iterations. 
Usually, Vast scaling focuses on extremely stable variables but this effect is mitigated as the transient behavior of the considered database for the analysis increases.

\begin{figure}
	\centering
	\includegraphics[width = \linewidth]{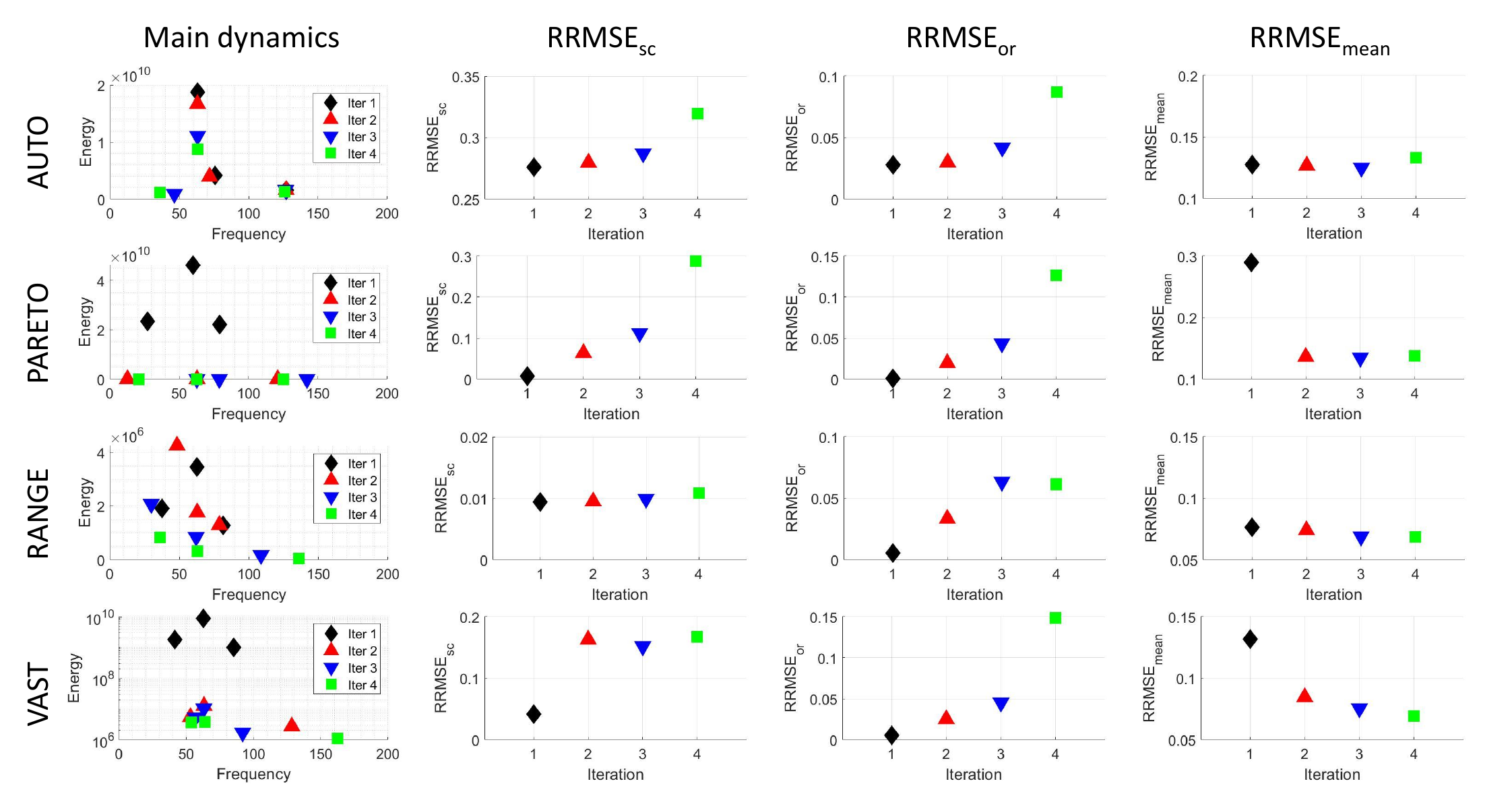}
	\caption{From left to right: Three main frequencies calculated as in Equation \ref{e42}, $RRMSE_{sc}$, $RRMSE_{or}$ and $RRMSE_{mean}$. From top to bottom: auto scaling, range scaling, pareto scaling and vast scaling. The analysis corresponds to Database 2, i.e., to the simulation obtained by imposing a sinusoidal perturbation having frequency $f = 10$ Hz and amplitude A = 0.25, and carried out with a detailed kinetic mechanism accounting for 53 species. \label{fig:Results2}}
\end{figure}

\begin{figure}
	\centering
	\includegraphics[width = \linewidth]{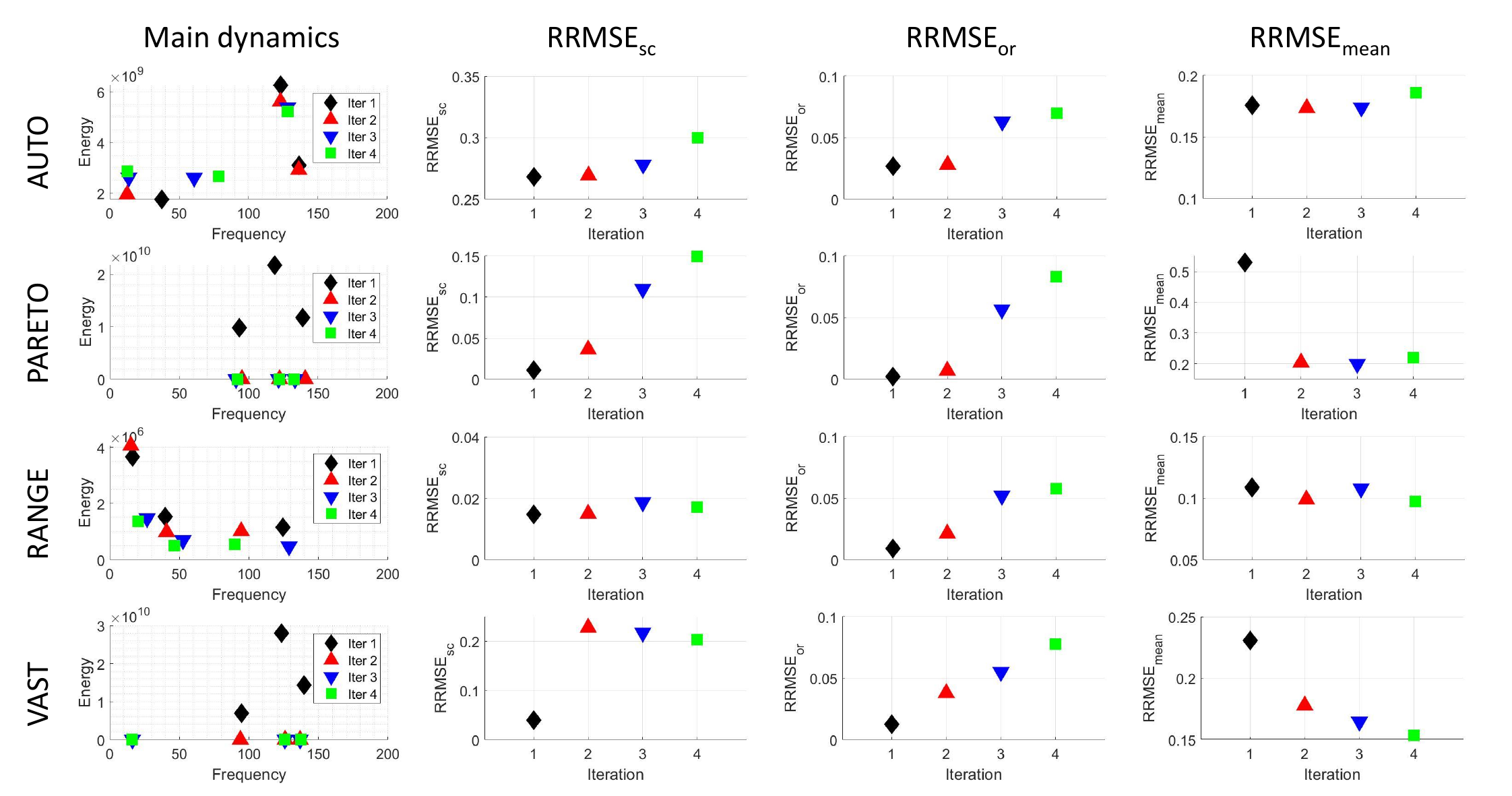}
	\caption{From left to right: Three main frequencies calculated as in Equation \ref{e42}, $RRMSE_{sc}$, $RRMSE_{or}$ and $RRMSE_{mean}$. From top to bottom: auto scaling, range scaling, pareto scaling and vast scaling. The analysis corresponds to Database 3, i.e., to the simulation obtained by imposing a sinusoidal perturbation having frequency $f = 20$ Hz and amplitude A = 0.25, and carried out with a detailed kinetic mechanism accounting for 53 species. \label{fig:Results3}}
\end{figure}

When databases 2 and 3 are analyzed (Figures \ref{fig:Results2} and \ref{fig:Results3}, respectively), similar results are obtained. 
For the main dynamics, Auto scaling provides similar frequencies for all the iterations, Pareto and Vast scaling have different dynamics for the first iteration and find same frequencies for the rest; while Range finds different frequencies for each iteration. 
With regard to the different reconstruction error metrics, the behavior of each scaling method follows the same behavior observed for the first database. 
The only metric that changes is the  $RRMSE_{mean}$ for Vast scaling, which behaves similarly to Pareto scaling. 
This is mainly due to the fact transient behavior for the numerical simulations 2 and 3 does not dominate as more cycles are taken into account. 

Summarizing the results for the three databases, Auto scaling provides the worst reconstruction error, although the clusters have similar dynamics. 
Range scaling gives the lowest $RRMSE_{mean}$ on the last iteration, while each cluster of variables on this scaling method have different main dynamics. 
Pareto behaves similarly to Auto scaling from the second iteration onward, although it allows for a better reconstruction of the first cluster. 
When Vast scaling is adopted, the first cluster appears to find different dynamics with respect to the rest of iterations and a great improvement for $RRMSE_{mean}$ is observed from second iteration onwards. 

After assessing the contribution for the different scaling methods, the new iterative formulation of the algorithm was compared with the standard HODMD. 
To accomplish this task, the reconstruction error relative improvement ($\eta$) between the two different algorithms is analyzed, which is defined as follows

\begin{equation}
	\eta = \dfrac{RRMSE_{HODMD}-RRMSE_{h-HODMD}}{RRMSE_{HODMD}},\label{eq:Imp}
\end{equation}

Equation \ref{eq:Imp} indicates that if a given variable is better reconstructed by h-HODMD than by the standard HODMD, a positive value for the improvement is obtained, wherease the opposite occurs when the aforementioned quantity is negative. 

\begin{figure}
	\centering
	\includegraphics[width = \linewidth]{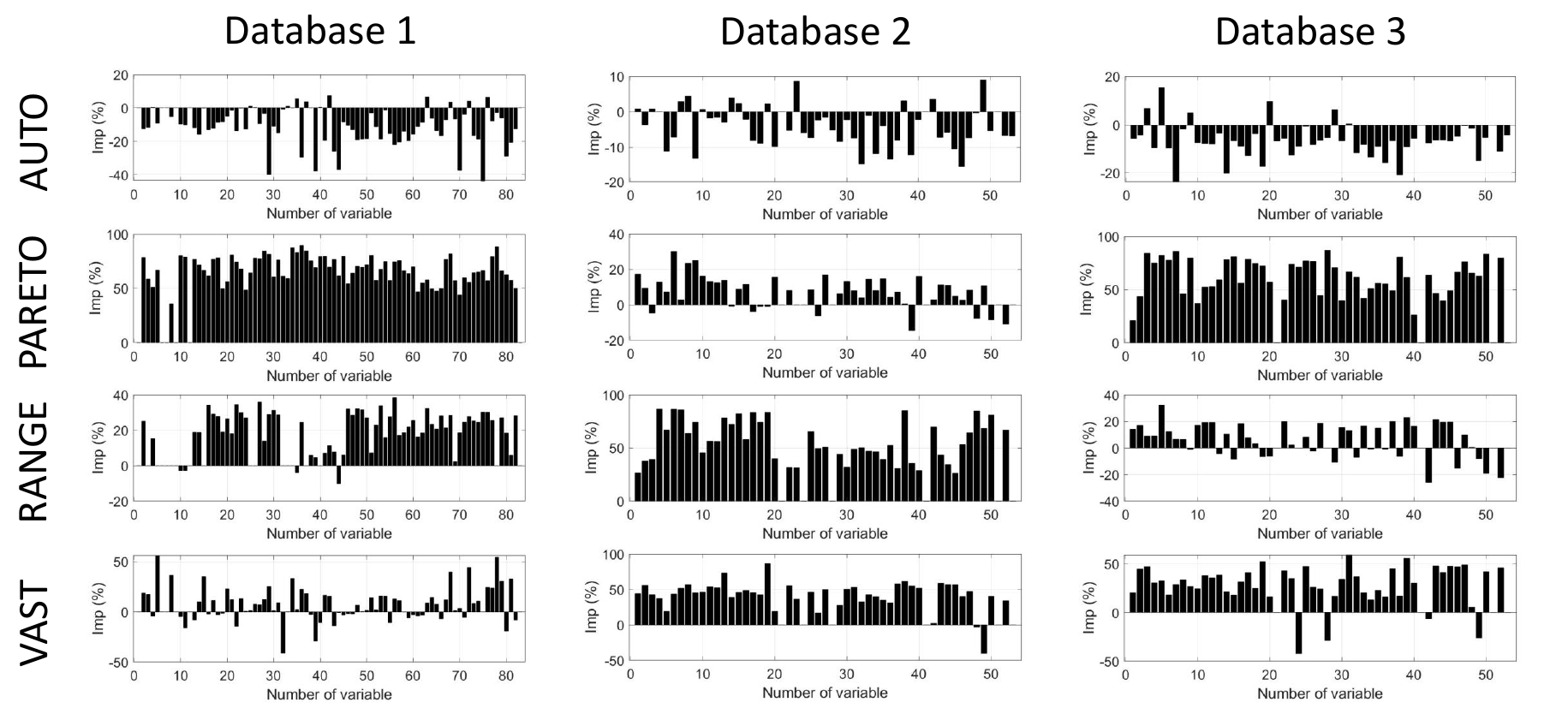}
	\caption{Reconstruction error improvement for all the variables of the chemical system computed as reported in Equation \ref{eq:Imp} for Database 1 (simulation obtained by imposing a sinusoidal perturbation having frequency $f = 10$ Hz and amplitude A = 0.25, and carried out with a detailed kinetic mechanism accounting for 82 species), Database 2 (simulation obtained by imposing a sinusoidal perturbation having frequency $f = 10$ Hz and amplitude A = 0.25, and carried out with a detailed kinetic mechanism accounting for 53 species), Database 3 (simulation obtained by imposing a sinusoidal perturbation having frequency $f = 20$ Hz and amplitude A = 0.25, and carried out with a detailed kinetic mechanism accounting for 53 species), and considering different scaling criteria (from top to bottom: Auto, Range, Pareto, Vast). \label{fig:Improvement}}
\end{figure}

The improvement factor is shown in Figure \ref{fig:Improvement} for all the databases and scaling methods. 
By adopting Auto scaling, the worst results are achieved as the reconstruction error gets worse for most of the species. 
The use of Range scaling improves instead the reconstruction error for most of the variables, and the same happens for Pareto scaling.
Thus, these two scaling methods appear to be a better choice when h-HODMD is adopted. Lastly, Vast scaling appears to favor the iterative algorithm as the transient behavior decreases (e.g., as when moving from database 2 to database 3). 

\subsection{HODMD feature selection \label{sec:selection}}

As already discussed in Section \ref{sec:Meth}, the new iterative formulation for the HODMD is capable of achieving a lower reconstruction error by grouping different variables at each iteration, i.e., by applying a model reduction considering clusters of variables instead of the full input tensor. 
\par
\begin{table}
	\centering
	\begin{tabular}{|c|c|c|c|c|}
		\hline
		Iteration 1 & Iteration 2 & Iteration 3 & Iteration 5 & Iteration 7\\
		\hline
		$CH_4$ & $CO$ & $HO_2$ & $CH_3$ & $H_2O_2$ \\
		\hline
		$CO_2$ & $N_2O$ & $O$ & &\\
		\hline
		$H_2O$ & $NO$ & $OH$  &  &  \\
		\hline
		$N_2$ & $NO_2$& &  &\\
		\hline
		$O_2$ &  &  &   &\\
		\hline
		$T$ & &  &    &\\
		\hline
	\end{tabular}
	\caption{Clusters of variables obtained using h-HODMD for Database 3 (i.e., for the simulation obtained by imposing a sinusoidal perturbation having frequency $f = 20$ Hz and amplitude A = 0.25, and carried out with a detailed kinetic mechanism accounting for 53 species) and using range scaling for the preprocessing step. \label{tab:third_range}}
\end{table}

The groups of variables that are found by the iterative algorithm for the third database considering Range scaling are listed in Table \ref{tab:third_range}, while all the other cases (in terms of databases and scaling criteria) are made available to the reader in the supplemental material. 
In this work, only the most relevant species from a physical point of view were taken into account to perform the variables' clusters analysis, therefore in Table \ref{tab:third_range} only temperature, fuel, oxidizer, main oxidation products, main initiation and propagation radicals, and main nitrogen oxides are reported.
The tolerance $\epsilon_{v}$ was set equal to  $\epsilon_{v} = \textit{tol} \times iter$, where the value for \textit{tol} was set to be equal to the RRMSE of the sixth best reconstructed variable, so as the first cluster should always contain the main variables for the reacting system (i.e., temperature, fuel, oxidizer, and main oxidation products such as water and carbon dioxide).
If an iteration of the algorithm does not report any variable is then removed and the algorithm runs the next iteration until there are no variables left, so in this analysis the number of iterations is not fixed.
\par
Pareto scaling always fails at identifying groups of variables which are coherent from a chemical point of view, as seen in Table \ref{tab:third_range} and in Tables S{\MakeUppercase{\romannumeral 1}} to S{\MakeUppercase{\romannumeral 7}} of supplemental material. 
In fact, O and OH are placed in different clusters, and they are grouped instead with nitrogen oxides which have a much slower dynamics.
In addition, it also requires the maximum number of iterations to identify and to cluster all the selected variables. 
On the other hand, the application of Vast, and Range scaling yields to better results not only from a reconstruction error point of view, but also because of the possibility to have physics-based clusters consisting of species whose kinetic behavior is similar. 
In fact, the aforementioned scaling criteria result in two different clusters separating the species having slow dynamics (such as the nitrogen oxides) from fast dynamics (radicals responsible for propagation and branching reactions such as the oxygen radicals, as well as the ones involved in initiation reactions such as the methane radical). 
As shown in Table \ref{tab:third_range}, the three main nitrogen oxides (NO, NO$_{2}$, and N$_{2}$O) are grouped together when Range scaling is utilized, while another cluster is identified for oxy- and peroxy radicals (O, OH, and HO$_{2}$).
Also, this scaling only needs seven iterations to get this result, while nine iterations are needed when Vast scaling is employed.
\par
\begin{figure}
	\centering
	\includegraphics[width = \linewidth]{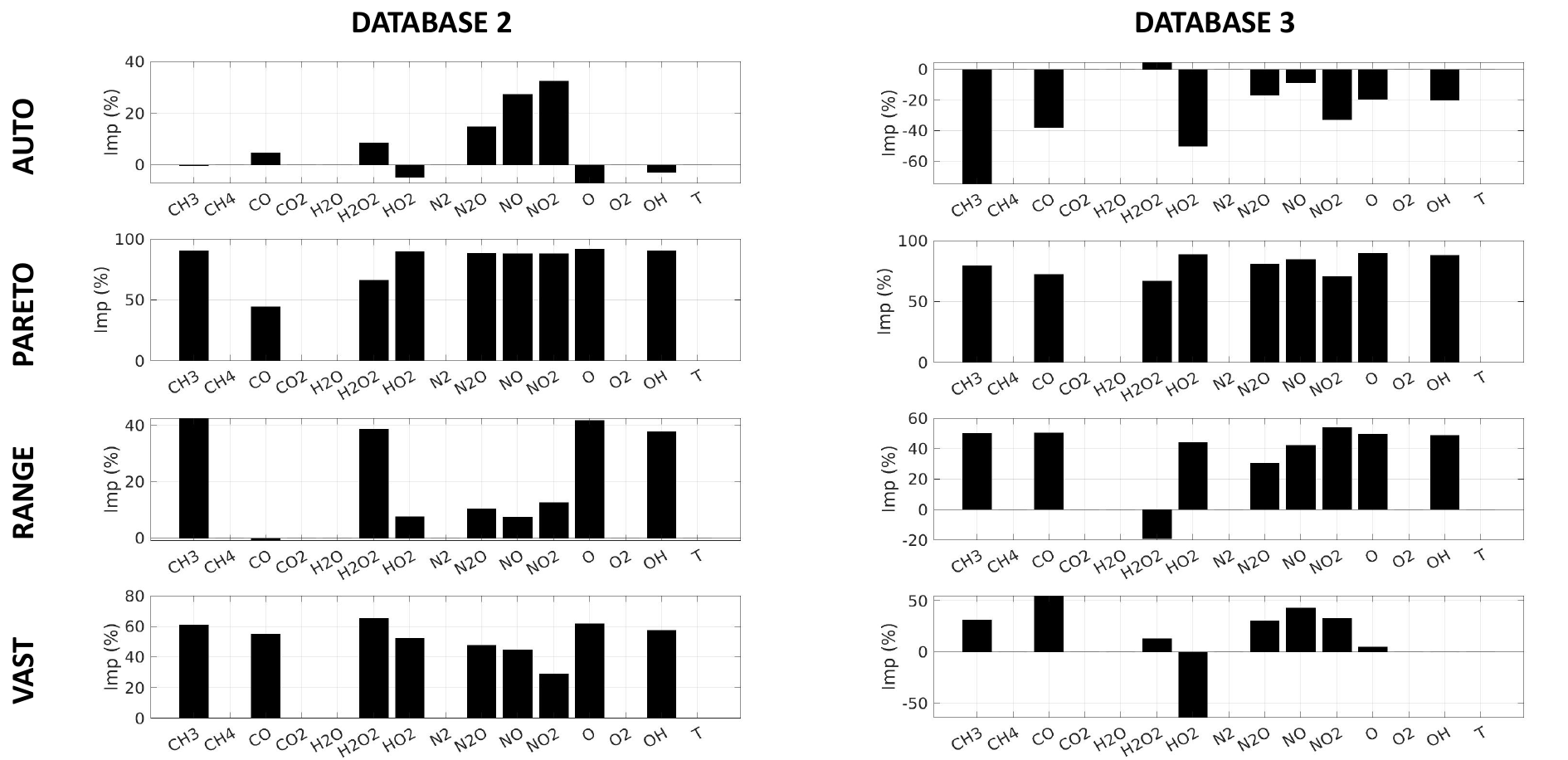}
	\caption{Reconstruction error improvement for the main variables of the chemical system computed as reported in Equation \ref{eq:Imp} for Database 2 (simulation obtained by imposing a sinusoidal perturbation having frequency $f = 10$ Hz and amplitude A = 0.25, and carried out with a detailed kinetic mechanism accounting for 53 species), and Database 3 (simulation obtained by imposing a sinusoidal perturbation having frequency $f = 20$ Hz and amplitude A = 0.25, and carried out with a detailed kinetic mechanism accounting for 53 species), considering different scaling criteria (from top to bottom: Auto, Range, Pareto, Vast). \label{fig:Improvement_relevant}}
\end{figure}

\begin{figure}
	\centering
	\includegraphics[width = 0.22\linewidth]{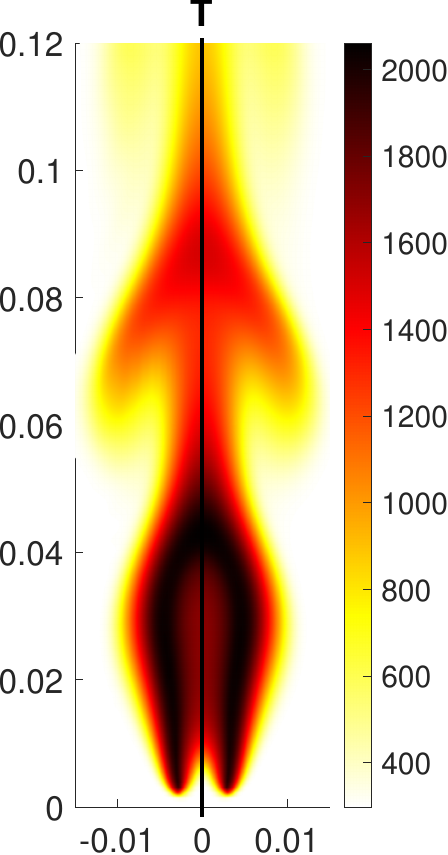} 
	\includegraphics[width = 0.22\linewidth]{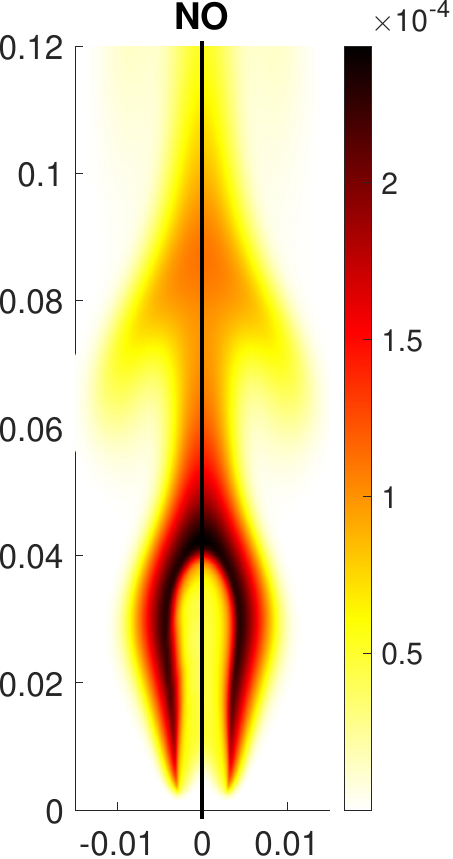} 
	\includegraphics[width = 0.22\linewidth]{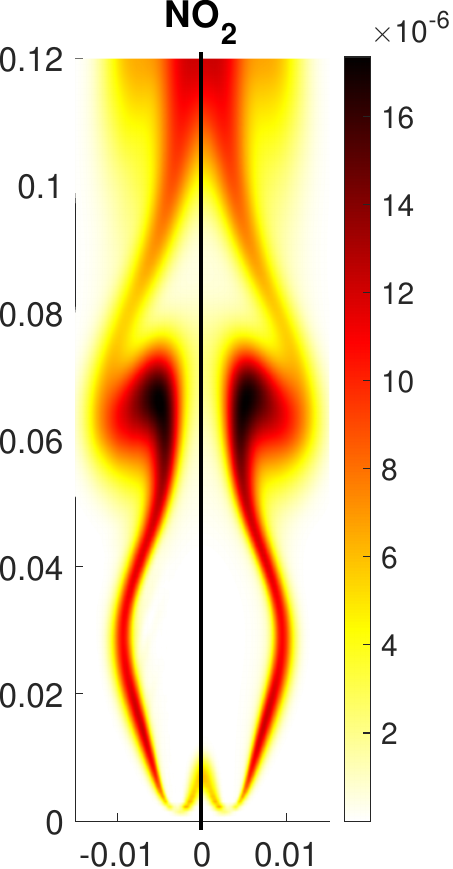} 
	\includegraphics[width = 0.22\linewidth]{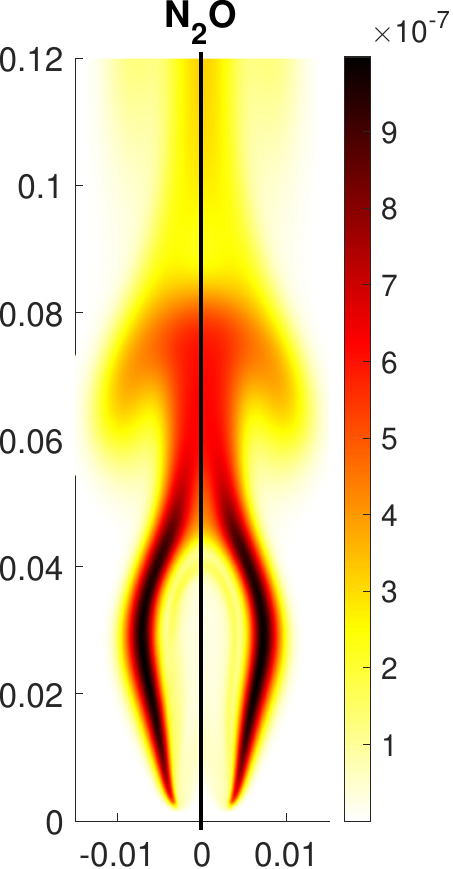} 
	\caption{Reconstruction of a representative snapshot of the flame for different variables using h-HODMD (Range scaling). The right side of each contour represents the original snapshot, while the left side of each contour represents the reconstruction with h-HODMD. Figures from left to right: contours of temperature, contours of NO, contours of NO$_{2}$, contours of N$_{2}$O.\label{fig:Rec}}
\end{figure}

As can be observed from Figure \ref{fig:Improvement_relevant}, Range scaling allows for a significant improvement ($\sim$40\%) in the reduced model's accuracy for the description of radicals that are fundamental for the reacting layer dynamics and liftoff height prediction such as OH, O, H$_{2}$O$_{2}$, and CH$_{3}$. 
Also from a qualitative comparison, such as the one reported in Figure \ref{fig:Rec} where the original (left side of each figure) and the reconstructed (right side of each figure) variables' profiles are compared, it is clear that the error observed when the proposed algorithm is adopted is negligible. 

\begin{figure}
	\centering
	\includegraphics[width = 0.22\linewidth]{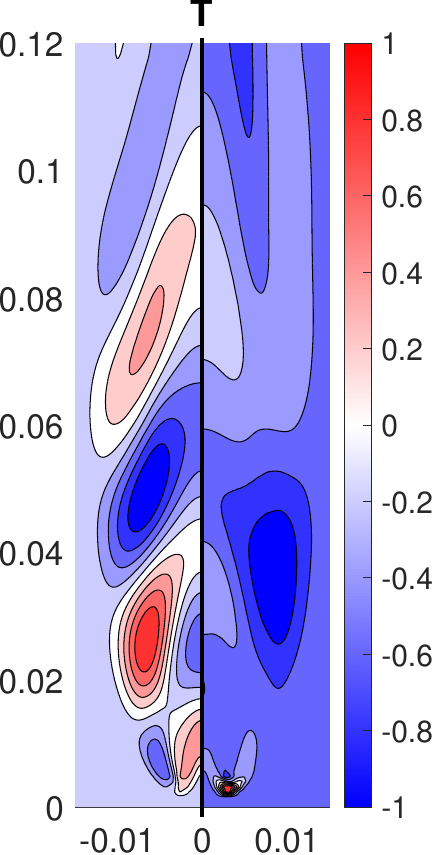}
	\includegraphics[width = 0.22\linewidth]{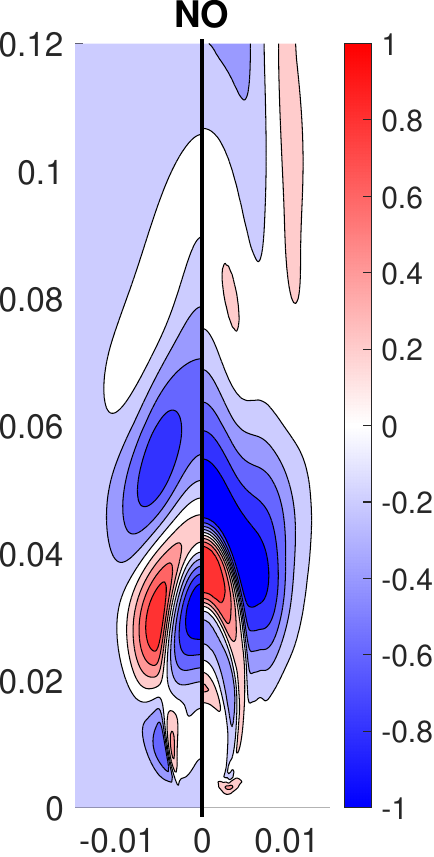}
	\includegraphics[width = 0.22\linewidth]{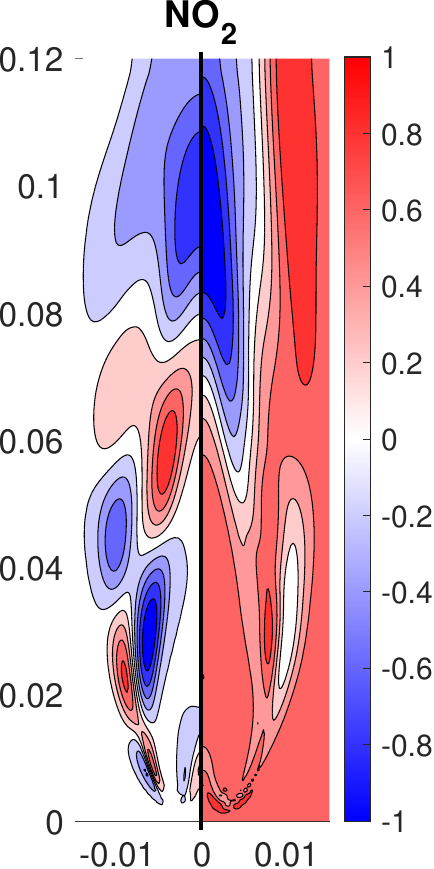}
	\includegraphics[width = 0.22\linewidth]{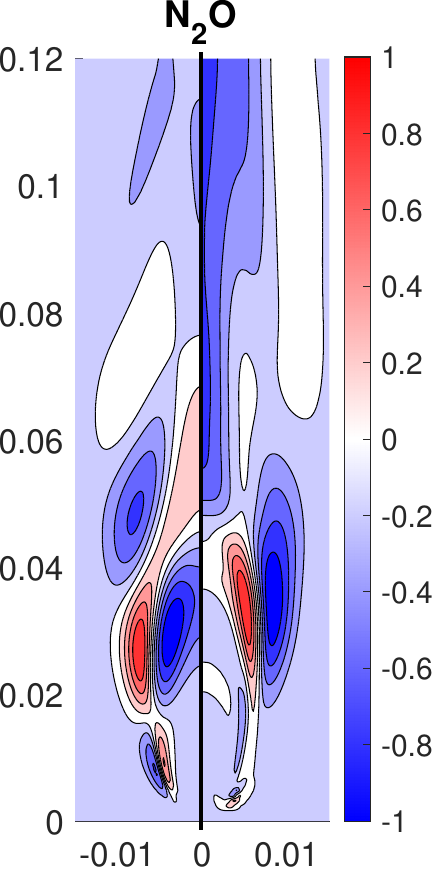}
	\caption{Real part of the two most important DMD modes (\textbf{left:} $\omega \sim 126$,\textbf{right:} $\omega \sim 25$) for some of the most relevant species using h-HODMD with range scaling criteria. The modes have been normalised between $-1$ and $1$.\label{fig:Modes}}
\end{figure}

Finally, the two most relevant modes for temperature and nitrogen oxides are reported in Figure \ref{fig:Modes}. 
In particular, the left side of each figure is representative for $\omega_1 \sim 126$, which is the frequency of perturbation of the simulation ($f_1 = \omega_1/2\pi \simeq 20$). This mode has, for all the species, a dominant component along the streamwise direction due to the effects of the sinusoidal velocity perturbation. 
Also, the modes for the nitrogen oxides are non-negative in the regions where these chemical species are formed and have the maximum concentration (i.e., in the regions where the reaction terms are stronger) while negative values are observed where the transport term is dominant and no reaction is occurring.
On the right side of each contour in Figure \ref{fig:Modes}, the frequency with highest energy $\omega_2 = 25$ is shown. This particular frequency was selected as it is one sub-harmonic of the main frequency ($\omega_2= \omega_1/5$) thus it is expected to have larger structures. 
For all the considered variables, it is possible to observe that the mode is focused on the large flow structures upstream and downstream, and it also has a peak in the liftoff and the flame border regions which are due to the strong gradients (both in the streamwise and in the spanwise directions) they undergo.

\section{Conclusions \label{sec:conclusions}}

The article introduces a new iterative algorithm based on HODMD, the hierarchical HODMD (h-HODMD), to formulate reduced order models for reacting flows.
The novelty in the proposed algorithm is represented by the possibility to perform model reduction by selecting groups of variables (clusters) thus achieving a reduction in the error observed for the reduced model. 
In addition, for each cluster it is possible to identify a main dynamics which is characteristic for the considered group of variables. 
This technique results to be particularly effective for the analysis of reacting flows, as a high number of variables is usually considered in the corresponding CFD simulations.
As the algorithm relies on modes with their frequencies, it can be adapted to decouple  diff dynamics in combustion systems. The h-HODMD iteratively identifies and isolates the best reconstructed variables by computing the RRMSE of each variable and by imposing a new tolerance settled by the user. 
Consequently, in addition to reduced order modeling, it can also be used for feature selection and variables' clustering tasks.

The new algorithm has been successfully applied in all the simulations and scaling methods analysed, obtaining a cluster at each iteration with an associated main dynamics. 
Auto scaling makes all the clusters to have the same dynamics, while it is not able to improve the reconstruction with standard HODMD.
The first cluster using Pareto scaling has different frequencies with respect to the rest, while greatly improving the reconstruction error, compared with HODMD.
Vast scaling has similar behavior on main dynamics as Pareto, while only making good improvements on the second and third database.
Finally, Range scaling makes all the clusters have different dynamics and makes improvement for most of the variables.

When analysing the clusters obtained with the algorithm, Pareto shows poor results and it is not able to identify groups of variables which are coherent from a chemical point of view. On the other hand, both Vast and Range scaling provide good results for both clustering and reconstructing the solution, while Range scaling gives better results overall.

This new iterative algorithm for higher order dynamic mode decomposition has been proved to be promising to study high-dimensional data such as the ones obtained from reacting flow simulation carried out with detailed kinetic mechanisms. In particular, the proposed method for feature selection and clustering of variables can be useful for the development of new reduced order models based on the physics knowledge. This allows new strategies to face increasingly large databases, reducing their associated computational cost.

\begin{acknowledgments}
A.C. and S.L.C. acknowledge the grant PID2020-114173RB-I00 funded by MCIN/AEI/ 10.13039/501100011033. S.L.C. and A.C. acknowledge the support of Comunidad de Madrid through the call Research Grants for Young Investigators from Universidad Politécnica de Madrid. 
The authors gratefully acknowledge the Universidad Politécnica de Madrid (www.upm.es) for providing computing resources on Magerit Supercomputer.
A.C. also acknowledges the support of Universidad Politécnica de Madrid, under the programme ‘Programa Propio’. 
G.D.A. has received funding from the Fonds National de la Recherche Scientifique (FRS-FNRS) through a FRIA fellowship. 
A.P. acknowledges funding from the European Research Council (ERC) under the European Union’s Horizon 2020 research and innovation programme under grant agreement No 714605.
\end{acknowledgments}

\bibliography{sample.bib}

\end{document}